\documentstyle[prb,aps,twocolumn,epsfig]{revtex}
\newcommand{\be}{\begin{eqnarray}}
\newcommand{\ee}{\end{eqnarray}}

\begin{document}
\draft
\twocolumn[\hsize\textwidth\columnwidth\hsize\csname @twocolumnfalse\endcsname

\title{Magnetic Field Dependent Optical Studies of a Layered Antiferromagnet Pr$%
_{1/2}$Sr$_{1/2}$MnO$_3$}
\author{J. H. Jung, H. J. Lee, and T. W. Noh\cite{email1}}
\address{Department of Physics and Center for Strongly Correlated Materials Research, 
\\
Seoul National University, Seoul 151-742, Korea}
\author{Y. Moritomo}
\address{CIRSE, Nagoya University, Nagoya 464-8603 and PRESTO, JST, Japan}
\author{Y. J. Wang and X. Wei}
\address{National High Magnetic Field Laboratory at Florida State University\\
Tallahassee, FL 32310, USA}
\date{\today }
\maketitle

\begin{abstract}
Magnetic field ($H$)-dependent optical conductivity spectra $\sigma (\omega
) $ and dielectric constant spectra $\varepsilon (\omega )$ of a layered
antiferromagnet ($A$-type) Pr$_{1/2}$Sr$_{1/2}$MnO$_3$ were presented. At
0.0 T in 4.2 K, the $\sigma (\omega )$ showed a small Drude-like peak in the
far-infrared region and a broad absorption peak in the mid-infrared region.
With increasing $H$, large spectral weights were transferred from high to
low energy regions with increase of the Drude-like peak. We found that a
polaron picture could explain most of the $H$-dependent spectral changes. By
comparing with the $H$-dependent spectral weight changes and $\varepsilon
(\omega )$ of Nd$_{1/2}$Sr$_{1/2}$MnO$_3$, which has the $CE$-type spin
ordering, we showed that a dimensional crossover from a 2-dimensional to a
3-dimensional metal occurs in Pr$_{1/2}$Sr$_{1/2}$MnO$_3$.
\end{abstract}

\pacs{PACS number; 75.50.Cc, 72.15.Gd, 75.30.Kz, 78.20.Ci}


\vskip1pc] \newpage

Doped manganites with chemical formula {\it R}$_{1-x}${\it A}$_x$MnO$_3$ (%
{\it R }= La, Pr, Nd and {\it A} = Ca, Sr, Ba) have attracted much attention
due to their exotic electrical and magnetic properties. Basic physics of
doped manganites have been explained by the double exchange model based on
the strong Hund coupling between itinerant {\it e}$_g$ and localized {\it t}$%
_{2g}$ electrons.\cite{zener} However, additional degrees of freedom, such
as a polaron due to the Jahn-Teller distortion of the Mn$^{3+}$ ion\cite
{millis,roder} and an orbital fluctuation,\cite{maekawa,horsch} have been
suggested to explain colossal magnetoresistance\cite{jin} and charge/orbital
ordering\cite{cheong} phenomena.

In addition to ferromagnetic metallic ({\it x }$\sim $ 1/3) and charge
ordered insulating ({\it x }$\sim $ 1/2) ground states, another intriguing
ground states were reported in a high doping region (i.e., {\it x }$\geq $
1/2). It was proposed that some doped manganites, such as Nd$_{0.45}$Sr$%
_{0.55}$MnO$_3$ and Pr$_{1/2}$Sr$_{1/2}$MnO$_3$, have metallic ground states
where spins in a layer are ferromagnetically ordered but the coupling
between the layers is antiferromagnetically ordered.\cite{kawano} This spin
configuration is called the ''$A$-type'' antiferromagnetic ordering. The
realization of the layered antiferromagnetic metallic ground states was
explained by a kinetic energy gain when the {\it d}$_{x^2-y^2}$ orbitals
form a 2-dimensional (2D) band.\cite{akimoto} This intriguing ground state
was experimentally demonstrated in a single domain Nd$_{0.45}$Sr$_{0.55}$MnO$%
_3$ crystal by measuring anisotropic electric properties: a metallic
conduction within the ferromagnetic layers and an insulating behavior along
the antiferromagnetic directions below the N\'{e}el temperature.\cite
{kuwahara} Under a high magnetic field ($H$), these manganites in the $A$%
-type antiferromagnetic ground state experience large resistivity changes
with a strong hysteresis.\cite{tomioka}

In spite of these fascinating physical phenomena, optical investigations on
the layered antiferromagnetic materials have been rare. Recently, we
reported temperature dependent optical conductivity spectra $\sigma (\omega
) $ of Pr$_{1/2}$Sr$_{1/2}$MnO$_3$.\cite{jung_psmo} We demonstrated that the
polaron scenario should be better to explain the temperature dependent
optical spectra of Pr$_{1/2}$Sr$_{1/2}$MnO$_3$ than the orbital fluctuation
scenario. We also suggested that some observed features of $\sigma (\omega )$
could be explained by a dimensional crossover from a 3D to a 2D metal near
the N\'{e}el temperature. [Note that the direct transport data which
demonstrates Pr$_{1/2}$Sr$_{1/2}$MnO$_3$ has a 2D metallic state are still
lacking due to the unavailability of a single domain crystal.]

In this paper, we will report $H$-dependent $\sigma (\omega )$ of Pr$_{1/2}$%
Sr$_{1/2}$MnO$_3$. At 0.0 T in 4.2 K, $\sigma (\omega )$ show a Drude-like
peak in the far-infrared region and a broad absorption peak in the
mid-infrared region. With increasing $H$, the Drude-like peak increases and
the large spectral weight changes are observed below 4.0 eV. These spectral
weight changes can be explained within the polaron picture. Moreover, the $H$%
-dependences of the spectral weights and dielectric constant spectra $%
\varepsilon (\omega )$ of Pr$_{1/2}$Sr$_{1/2}$MnO$_3$ are different from
those of Nd$_{1/2}$Sr$_{1/2}$MnO$_3$, which has the $CE$-type
antiferromagnetic spin ordering. These experimental observations provide
further supports for the occurrence of the dimensional crossover in Pr$%
_{1/2} $Sr$_{1/2}$MnO$_3$.

A Pr$_{1/2}$Sr$_{1/2}$MnO$_3$ single crystal was grown by the floating zone
method. Details of sample growth and characterization were reported
elsewhere.\cite{moritomo97} Its $H$-dependent {\it dc} resistivity was
measured by the conventional four-probe method using a 20 T superconducting
magnet. For optical measurements, the crystal was polished up to 0.3 $\mu $m
using diamond paste. To remove surface damages due to the polishing process,
we carefully annealed the sample again in an O$_2$ atmosphere at 1000 $^o$C
just before optical measurements. The $H$-dependent reflectivity in the
energy region between 5.0 meV and 4.0 eV were measured using the
spectrophotometers at National High Magnetic Field Laboratory. [All of the
optical spectra reported in this paper were measured at 4.2 K.] Using the
Kramers-Kronig transformation, $\sigma (\omega )$ and $\varepsilon (\omega )$
were obtained. For the high frequency extrapolations, the room temperature
data between 4.0 and 30 eV were used.

\begin{figure}[tbp]
\epsfig{file=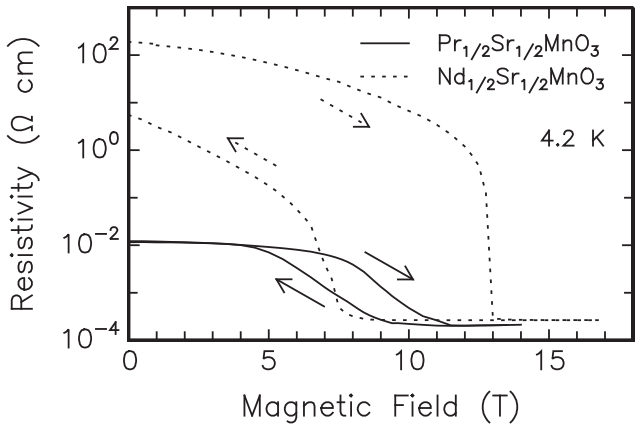,width=3.3in,clip=}
\vspace{2mm}
\caption{$H$-dependent {\it dc} resistivity of Pr$_{1/2}$Sr$_{1/2}$MnO$_3$
and Nd$_{1/2}$Sr$_{1/2}$MnO$_3$. }
\label{Fig:1}
\end{figure}

The solid lines in Fig. 1 show the $H$-dependent {\it dc} resistivity curves
of Pr$_{1/2}$Sr$_{1/2}$MnO$_3$, which were taken at $T$ $=$ 4.2 K. With
increasing $H$, the {\it dc} resistivity value ($\sim $ 10$^{-2}$ $\Omega $
cm at 0.0 T) is nearly constant up to 7.0 T, and then it starts to decrease
above 7.0 T. The {\it dc} resistivity value does not change at all above
11.5 T within our experimental error. With decreasing $H$, the {\it dc}
resistivity becomes nearly constant down to 9.5 T and then starts to
increase. Note that the {\it dc} resistivity shows a strong hysteresis.

For comparison, the $H$-dependent {\it dc} resistivity curves of Nd$_{1/2}$Sr%
$_{1/2}$MnO$_3$ are also shown as dotted lines in Fig. 1. Compared to the Pr$%
_{1/2}$Sr$_{1/2}$MnO$_3$ case, the $H$-dependent {\it dc} resistivity change
of Nd$_{1/2}$Sr$_{1/2}$MnO$_3$ is quite large. With increasing $H$, the high
resistivity value ($\sim $ 10$^2$ $\Omega $ cm at 0.0 T) is sharply
decreased near 13.0 T and also shows a very strong hysteresis. Note that the 
{\it dc} resistivity value at 0.0 T in Nd$_{1/2}$Sr$_{1/2}$MnO$_3$ in the $H$%
-decreasing run is quite smaller than that in the $H$-increasing run.

In Fig. 2(a), we show the $H$-dependent $\sigma (\omega )$ of Pr$_{1/2}$Sr$%
_{1/2}$MnO$_3$ during the $H$-increasing run. At 0.0 T, there are broad
peaks around 1.0 and 4.0 eV. As $H$ increases, the spectral weights near 1.0
eV and 3.0 eV are transferred to a lower energy region. The gap-like feature
in $\sigma (\omega )$ at 0.0 T changes into a Drude-like peak above 10.0 T.
The low frequency details of transferred spectral weights below 0.1 eV can
be seen in Fig. 2(b). At 0.0 T, there are sharp peaks due to the optic
phonon modes\cite{kim96} and a hint of small rise in $\sigma (\omega )$ in
the low frequency limit, suggesting the existence of the small Drude-like
peak. As $H$ increases, the phonon peaks become screened and the Drude-like
peak becomes more clear below 0.04 eV. The solid circle represents the {\it %
dc} conductivity value at 14.0 T. Therefore, $\sigma (\omega )$ below 0.1 eV
at the high $H$ should be viewed as two parts, i.e., the Drude and the large
incoherent absorption peaks.\cite{kim96}

\begin{figure}[tbp]
\epsfig{file=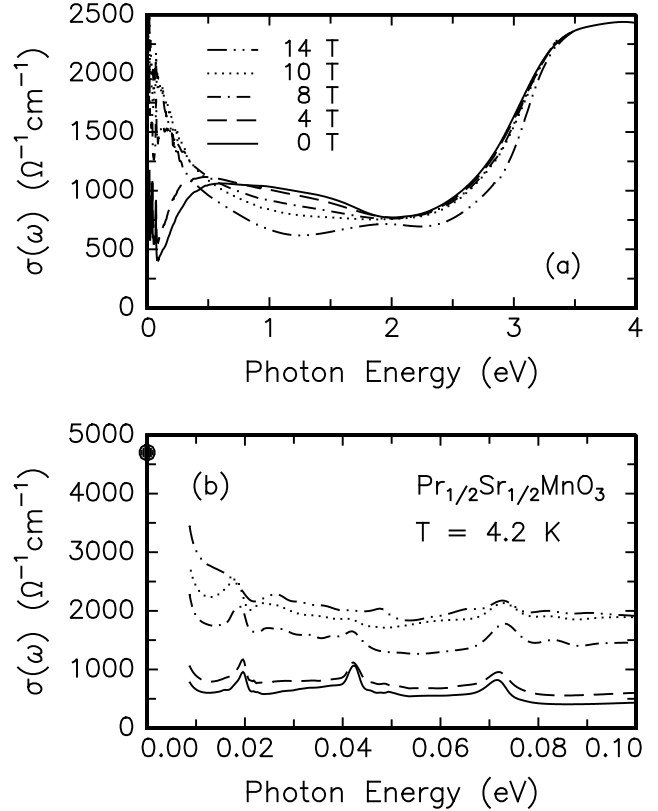,width=3.3in,clip=}
\vspace{2mm}
\caption{$\sigma (\omega )$ of Pr$_{1/2}$Sr$_{1/2}$MnO$_3$ below (a) 4.0 eV
and (b) 0.1 eV. In (b), the solid circle represents the {\it dc}
conductivity value at 14 T.}
\label{Fig:2}
\end{figure}

In our earlier studies on the temperature dependent $\sigma (\omega )$ of Pr$%
_{1/2}$Sr$_{1/2}$MnO$_3$,\cite{jung_psmo} we observed similar spectral
weight changes. In the high frequency region above 2.0 eV, there are two
contributions. Around 4.0 eV, there is a peak due to the charge transfer
transition from the O{\it \ }2{\it p} level to the Mn 3{\it d} level.\cite
{jung97} And, there is a broad peak due to the optical transition between
the Hund's rule split bands near 3.0 eV,\cite{moritomo_op} however it is
much weaker than the charge transfer transition. And, it was also shown that
the polaron scenario was better to describe the optical spectra in the low
frequency region than the orbital fluctuation scenario.

After subtracting the charge transfer transition contribution from the
measured $\sigma (\omega )$, we analyzed $H$-dependent $\sigma (\omega )$
below 2.0 eV within the polaron picture:

\begin{equation}
\sigma (\omega <\text{2.0 eV})=\sigma _D(\omega )+\sigma _I(\omega )+\sigma
_{II}(\omega )\text{,}
\end{equation}
where $\sigma _D(\omega )$, $\sigma _I(\omega )$, $\sigma _{II}(\omega )$
represent the contributions due to free carriers, incoherent polaron
absorption,\cite{kaplan,jung98} and the inter-orbital transition between the
Jahn-Teller split levels,\cite{jung98,allen} respectively. The simple Drude
model was used for $\sigma _D(\omega )$, and two Gaussian functions were
used for $\sigma _I(\omega )$ and $\sigma _{II}(\omega )$. Then, we
estimated the $H$-dependent optical strengths $S_D$, $S_I$, and $S_{II}$ by
integrating $\sigma _D(\omega )$, $\sigma _I(\omega )$, and $\sigma
_{II}(\omega )$, respectively. Details of this procedure were published
elsewhere.\cite{jung_psmo}

\begin{figure}[tbp]
\epsfig{file=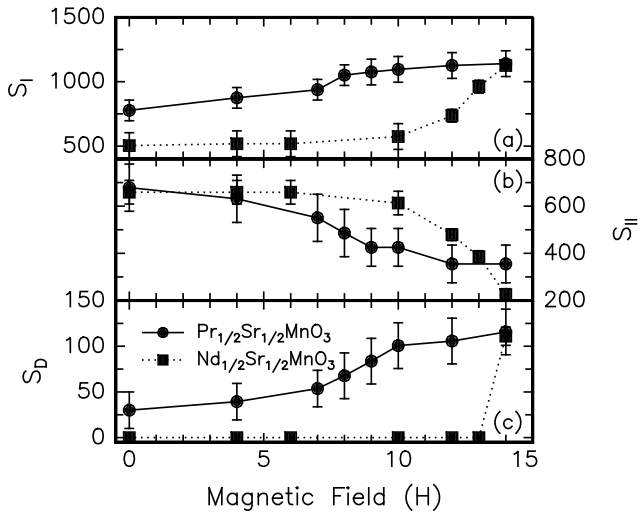,width=3.3in,clip=}
\vspace{2mm}
\caption{Optical strengths (a) $S_I$, (b) $S_{II}$, and (c) $S_D$ for Pr$%
_{1/2}$Sr$_{1/2}$MnO$_3$ (solid circles) and Nd$_{1/2}$Sr$_{1/2}$MnO$_3$
(solid squares). [All the units are $\Omega ^{-1}$cm$^{-1}$eV.]}
\label{Fig:3}
\end{figure}

Figures 3(a), (b), and (c) show the $H$-dependence of $S_I$, $S_{II}$, and $%
S_D$ for Pr$_{1/2}$Sr$_{1/2}$MnO$_3$ (solid circles) during the $H$%
-increasing run, respectively. [For comparison, Nd$_{1/2}$Sr$_{1/2}$MnO$_3$
data were also shown as solid squares.] With increasing $H$, $S_I$ and $S_D$
increase smoothly, but $S_{II}$ shows the opposite behavior. The $H$%
-dependences of $S_I$, $S_{II}$, and $S_D$ for Nd$_{1/2}$Sr$_{1/2}$MnO$_3$
are similar to those for Pr$_{1/2}$Sr$_{1/2}$MnO$_3$. And, at high $H$, the
values of $S_I$ and $S_D$ are nearly the same for both samples. However,
there are a few important differences between Pr$_{1/2}$Sr$_{1/2}$MnO$_3$
and Nd$_{1/2}$Sr$_{1/2}$MnO$_3$ data. First, at $H=$ 0.0 T, $S_D$ of Pr$%
_{1/2}$Sr$_{1/2}$MnO$_3$ remains as a finite value, while that of Nd$_{1/2}$%
Sr$_{1/2}$MnO$_3$ becomes nearly zero. Second, the $H$-dependences of $S_I$
and $S_D$ are much weaker for Pr$_{1/2}$Sr$_{1/2}$MnO$_3$. These $H$%
-dependences agree with Fig. 1, where {\it dc} resistivity of Pr$_{1/2}$Sr$%
_{1/2}$MnO$_3$ is smaller than that of Nd$_{1/2}$Sr$_{1/2}$MnO$_3$ by four
orders of magnitude at zero field but becomes nearly the same at high $H$.

Although there are no direct experimental evidences such as anisotropic
transport measurements on a single domain crystal, Pr$_{1/2}$Sr$_{1/2}$MnO$_3
$ is believed to have a 2D metallic behavior in the $A$-type spin ordered
states. On the other hand, the charge/orbital ordered state of Nd$_{1/2}$Sr$%
_{1/2}$MnO$_3$ with the $CE$-type ordering should be insulating.\cite
{kuwahara_science} The weaker $H$-dependences of $S_I$ and $S_D$ for Pr$%
_{1/2}$Sr$_{1/2}$MnO$_3$ can be easily understood. Moreover, $\sigma (\omega
)$ of Pr$_{1/2}$Sr$_{1/2}$MnO$_3$, shown in Fig. 2(b), show a weak
Drude-like peak even at $H=$ 0.0 T. To get more informations on the ground
state, we estimated the value of the total polaron absorption $S_{tot}$ ($=$ 
$S_I+$ $S_D$). It is found that the value of $S_{tot}$ at 0.0 T is by about
65 \% ($\sim $ 2/3) smaller than that at 14.0 T. The existence of the
Drude-like peak and the decrease of $S_{tot}$ by an amount of $\sim $ 1/3 at
zero field support that the ground state of Pr$_{1/2}$Sr$_{1/2}$MnO$_3$
should be a 2D metal.

\begin{figure}[tbp]
\epsfig{file=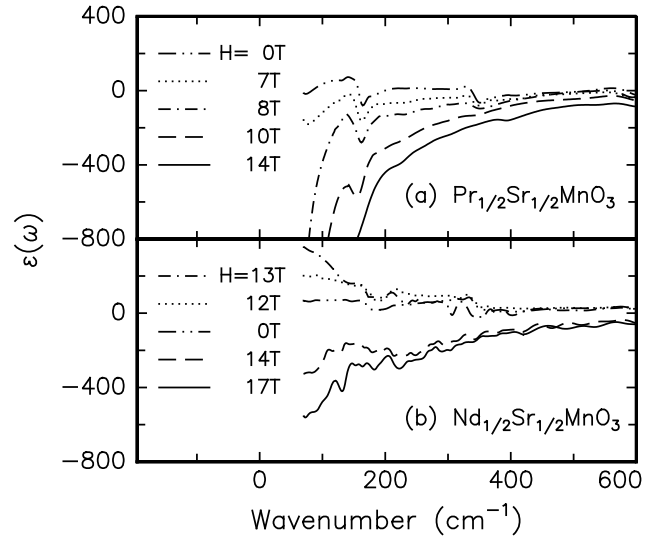,width=3.3in,clip=}
\vspace{2mm}
\caption{$H$-dependent $\varepsilon (\omega )$ of (a) Pr$_{1/2}$Sr$_{1/2}$MnO%
$_3$ and (b) Nd$_{1/2}$Sr$_{1/2}$MnO$_3$.}
\label{Fig:4}
\end{figure}

Further insights on Pr$_{1/2}$Sr$_{1/2}$MnO$_3$ can be obtained from the $H$%
-dependent $\varepsilon (\omega )$, shown in Fig. 4(a). At 0.0 T, $%
\varepsilon $ is nearly $\omega $-independent, but the value of $\varepsilon 
$ at the low frequency is less than zero. This result implies that Pr$_{1/2}$%
Sr$_{1/2}$MnO$_3$ should be a metal at 0.0 T, which is consistent with the
existence of the Drude-like peak in Fig. 2(b). With increasing $H$, $%
\varepsilon (\omega )$ decrease further and clearly show a metallic behavior
at 14.0 T. As a comparison, we also plot the $H$-dependent $\varepsilon
(\omega )$ for Nd$_{1/2}$Sr$_{1/2}$MnO$_3$ in Fig. 4(b). At 0.0 T, $%
\varepsilon $ is positive and nearly $\omega $-independent at all
frequencies, consistent with a typical insulator response.\cite{wooten} With
increasing $H$ up to 13.0 T, $\varepsilon (\omega )$ in the low frequency
region increases. With increasing $H$ further, $\varepsilon (\omega )$
suddenly become negative and finally show a typical metallic response at
17.0 T.\cite{wooten} In our earlier study for the $H$-dependent $\varepsilon
(\omega )$ of Nd$_{1/2}$Sr$_{1/2}$MnO$_3$,\cite{jung_melting} we
demonstrated that the divergence behavior of $\varepsilon (\omega )$ near
the insulator-metal transition boundary, i.e., 13.0 T, could be explained by
a dielectric anomaly near a percolation threshold. However, in Pr$_{1/2}$Sr$%
_{1/2}$MnO$_3$, such a dielectric anomaly of $\varepsilon (\omega )$ cannot
be observed near the 10.0 T region, where {\it dc} resistivity experiences a
large change.

The {\it dc} resistivity change near the 10.0 T region can be understood in
terms of a dimensional crossover from 2D to 3D metals. If the zero field
ground state of Pr$_{1/2}$Sr$_{1/2}$MnO$_3$ is a 2D metal, the sample might
be composed of multi-magnetic domains with their antiferromagnetic
directions randomly pointing along {\it x}-, {\it y}-, and {\it z}-
directions. For such sample, the percolating metallic paths should be
formed. That is the reason why we could observe the existence of the
Drude-like peak and the absence of the dielectric anomaly in Pr$_{1/2}$Sr$%
_{1/2}$MnO$_3$. In optical measurements, the light will probe the average
response of the multi-magnetic domains. Effectively, 2/3 of its response
comes from the metallic response within the ferrimagnetically ordered layer
and 1/3 comes from the insulating response along the antiferromagnetic
direction. Since the metallic response dominates in the low frequency
region, the value of $S_{tot}$ could be reduced approximately by 1/3 near
the dimensional crossover, in agreement with our experimental observation.
The detailed process of the dimensional crossover is not known yet. Using $%
^{55}$Mn NMR experiment, Allodi {\it et al}.\cite{allodi} suggested the
percolative origin for the ferromagnetic transition in Pr$_{1/2}$Sr$_{1/2}$%
MnO$_3$. Compared with the Nd$_{1/2}$Sr$_{1/2}$MnO$_3$ case, such a
percolative process is quite plausible. However, more studies are required
to understand details of the dimensional crossover.

In summary, we investigated the magnetic field dependent optical
conductivity spectra and dielectric constant spectra of an $A$-type spin
ordered half-doped manganite Pr$_{1/2}$Sr$_{1/2}$MnO$_3$. In comparison with
Nd$_{1/2}$Sr$_{1/2}$MnO$_3$, which show the $CE$-type spin ordering, we
showed that a dimensional crossover from a 2-dimensional to a 3-dimensional
metal should occur in Pr$_{1/2}$Sr$_{1/2}$MnO$_3$.

We acknowledge Dr. H. C. Kim and Dr. H.-C. Ri for help in magnetoresistance
measurements. This work was supported by Ministry of Science and Technology
through the Nanostructure Technology Project and by Ministry of Education
through the BK-21 Program. The work by Y. M. was supported by a Grant-In-Aid
for Science Research from the Ministry of Education, Science, Sports, and
Culture. Part of this work was performed at the National High Magnetic Field
Laboratory, which is supported by NSF Cooperative Agreement No. DMR-9016241
and by the State of Florida.

\end{document}